\documentclass[prl, preprint, nofootinbib, floatfix]{revtex4-1}
\usepackage{amsmath,amssymb,latexsym,mathrsfs}
\usepackage{graphicx,subfigure}
\usepackage{color}
\usepackage{hyperref}

\hypersetup{
    colorlinks=true,
    linkcolor=red,
    citecolor=blue,
}
\begin{document}
\title{Gauss-Bonnet inflation and swampland}
\author{Zhu Yi}
\email{yizhu92@hust.edu.cn}
\affiliation{School of Physics, Huazhong University of Science and Technology, Wuhan, Hubei 430074, China}
\author{Yungui Gong}
\email{yggong@hust.edu.cn}
\affiliation{School of Physics, Huazhong University of Science and Technology, Wuhan, Hubei 430074, China}

\begin{abstract}
The two swampland criteria are generically in tension with the single field slow-roll inflation because the first swampland criterion requires small tensor to scalar ratio while
the second swampland criterion requires large tensor to scalar ratio. The challenge to the single field slow-roll inflation imposed by the swampland criteria can be avoided
by modifying the relationship between the tensor to scalar ratio and the slow-roll parameter.
We show that the Gauss-Bonnet inflation with the coupling function inversely proportional to the potential overcomes the challenge
by adding a constant factor in the relationship between the tensor to scalar ratio and the slow-roll parameter.
For the Gauss-Bonnet inflation, while the swampland criteria are satisfied,
the slow-roll conditions are also fulfilled, so the scalar spectral tilt and the tensor to scalar ratio are consistent with the observations.
We use the potentials for chaotic inflation and the E-model as examples to show that the models pass all the constraints.
The swampland criteria may imply Gauss-Bonnet coupling.
\end{abstract}
\maketitle

\section{introduction}
Inflation solves the flatness and horizon problems in standard cosmology \cite{Guth:1980zm,Linde:1981mu,Albrecht:1982wi,Starobinsky:1980te,Sato:1980yn},
and is usually modelled by a single slow-roll scalar field which is obtained from low-energy Effective Filed Theories.
In order to embed such scalar fields in a quantum theory successfully, they have to satisfy two criteria \cite{Ooguri:2006in,Vafa:2005ui}. These two Swampland criteria are
\begin{itemize}
  \item Swampland Criterion I ($\mathcal{SC}$I) \cite{Ooguri:2016pdq}: The
scalar field excursion, normalized by the reduced planck mass, in field space is bounded from above
  \begin{equation}\label{swampland1}
     |\Delta \phi|  \leq d,
  \end{equation}
where the reduced planck mass $M_{pl}=1/\sqrt{8\pi G}=1$, and the order 1 constant $d\sim\mathcal{O}(1)$.
  \item Swampland  Criterion II ($\mathcal{SC}$II) \cite{Obied:2018sgi}: The
 gradient of the filed potential $V$ with $V>0$ should satisfy the lower bound
  \begin{equation}\label{swampland2}
    \frac{|V'|}{V}\geq c.
  \end{equation}
  where $V'=dV/d\phi$ and the order 1 constant $c\sim \mathcal{O}(1)$.
\end{itemize}
Obviously, the second criterion \eqref{swampland2} violates the slow-roll condition and poses a threat to the inflation model by
requiring a large tensor to scalar ratio $r\sim 8c^2$. Even if we chose $c=0.1$ \cite{Kehagias:2018uem},
it is still inconsistent with the observational constraint $r_{0.002}<0.064$ \cite{Akrami:2018odb,Ade:2018gkx} because $r\sim 0.08>0.064$.
As point out in Ref. \cite{Agrawal:2018own}, a viable way to solve this problem is by using the models with
the tensor to scalar ratio $r$ reduced  by a factor while keeping the lower bound on the field excursion $\Delta\phi$ by $r$ required by the Lyth bound \cite{Lyth:1996im},
such as the warm inflation \cite{Berera:1995ie}.
See Ref. \cite{Brennan:2017rbf,Das:2018hqy,Das:2018rpg,Motaharfar:2018zyb,Ashoorioon:2018sqb,Lin:2018kjm,Lin:2018rnx,Kinney:2018nny,
Achucarro:2018vey,Andriot:2018wzk,Park:2018fuj,Garg:2018reu,Garg:2018zdg,
Schimmrigk:2018gch,Dimopoulos:2018upl,
Matsui:2018bsy,Ben-Dayan:2018mhe,Brahma:2018hrd,Roupec:2018mbn,Blaback:2018hdo,
Odintsov:2018zai,Kawasaki:2018daf,Wang:2018kly} for the other papers about this issue.

In Ref. \cite{Yi:2018gse}, we find a powerful mechanism to reduce the tensor to scalar ratio $r$.
With the help of the Gauss-Bonnet term, for any potential,
the tensor to scalar $r$ is reduced by a factor $1-\lambda$ with the order 1 parameter $\lambda$, which may possibly  solve the swampland problem.
Generally, the predictions of the inflation, $n_s$ and $r$, are calculated under the slow-roll condition, $(V'/V)^2\ll1$.
If the second swampland criterion \eqref{swampland2} is satisfied, then the slow-roll condition will be violated,
and the predictions $n_s$ and $r$ may be unreliable. But in the case with Gauss-Bonnet coupling,
the  slow-roll condition is $(1-\lambda)(V'/V)^2\ll1$, so even the  second swampland criterion \eqref{swampland2} is satisfied,
as long as $1-\lambda\ll1$, the model still satisfies the slow-roll condition, so the slow-roll results are applicable.
In this paper, we show that with the help of the Gauss-Bonnet coupling, the inflation model satisfies not only the swampland criteria but also the observational constraints.

This paper is organized as follows. In Sec. II, we give a brief introduction to the Gauss-Bonnet inflation,
and point out the reason why it is easy to satisfy the swampland criteria for the model.
In Sec. III, we use the power-law potential and the E-model to show that all the constraints can be satisfied.
We conclude the paper in Sec. IV.

\section{The Gauss-Bonnet inflation}
\label{sec1}
The action for Gauss-Bonnet inflation is \cite{Rizos:1993rt,Kanti:1998jd,Nojiri:2005vv}
\begin{equation}\label{action1}
S=\frac{1}{2}\int\sqrt{-g}d^4 x \left[R-g^{\mu\nu}\partial_\mu\phi\partial_\nu\phi-2V(\phi)-\xi(\phi)R_{GB}^2\right],
\end{equation}
where $R^{2}_{\rm GB}
= R_{\mu\nu\rho\sigma} R^{\mu\nu\rho\sigma} - 4 R_{\mu\nu}
R^{\mu\nu} + R^2$ is the Gauss-Bonnet term which is a pure topological term in four dimensions,
and $\xi(\phi)$ is the Gauss-Bonnet coupling function. In this paper, we use \cite{vandeBruck:2016xvt,Yi:2018gse}
\begin{equation}
\label{xiphi}
    \xi(\phi)=\frac{3\lambda}{4 V(\phi)+\Lambda_0},
\end{equation}
where $0<\lambda<1$.
The parameter $\Lambda_0\ll (10^{16}\text{Gev})^4$ added here  is to avoid the reheating problem of Gauss-Bonnet inflation \cite{vandeBruck:2016xvt},
and it can be ignored during inflation, so in this paper we neglect the effect of $\Lambda_0$.
In terms of the horizon flow slow-roll parameters \cite{Schwarz:2001vv},
the slow-roll conditions are
\begin{gather}
\label{eps1}
  \epsilon_1=-\frac{\dot{H}}{H^2}\ll1,\quad \epsilon_2=\frac{\dot{\epsilon_1}}{H\epsilon_1}\ll1,
\end{gather}
the scalar spectral tilt $n_s$  and the tensor to scalar ratio $r$ are \cite{Yi:2018gse}
\begin{gather}
\label{ns1}
n_s-1= -2\epsilon_1-4\epsilon_2,\\
\label{r1}
r=16(1-\lambda)\epsilon_1.
\end{gather}
In terms of the potential, the slow-roll parameters are expressed as \cite{Guo:2010jr}
\begin{gather}\label{sl1}
\epsilon_1=\frac{1-\lambda}{2}\left(\frac{V'}{V}\right)^2, \quad \epsilon_2=-2(1-\lambda)\left[\frac{V''}{V}-\left(\frac{V'}{V}\right)^2\right].
\end{gather}
Due to the factor $1-\lambda$ in Eqs. \eqref{sl1}, even if the gradient of the potential is consistent with the second swampland criterion $\mathcal{SC}$II, $|V'|/V>c$,
the slow-roll conditions \eqref{eps1} can still be satisfied as long as $\lambda$ is close to 1.
So the slow-roll results \eqref{ns1}, \eqref{r1} and \eqref{sl1} are applicable to the case satisfying the second swampland criterion $\mathcal{SC}$II.
Substituting Eq. \eqref{sl1} into Eq. \eqref{r1}, we get
\begin{equation}
\label{r2}
  r=8(1-\lambda)^2\left( \frac{V'}{V}\right)^2.
\end{equation}
From Eq. \eqref{r2}, we see that while the  second swampland criterion $\mathcal{SC}$II is satisfied, the tensor to scalar ratio $r$ can still be very small, as long as $1-\lambda$ is small enough. 

Now we discuss the first swampland criterion $\mathcal{SC}$I for the field excursion. The Lyth bound tells us that \cite{Lyth:1996im,Yi:2018gse}
\begin{equation}\label{lythb}
  \Delta \phi>\Delta N\sqrt{\frac{r}{8}}=(1-\lambda)\Delta N \frac{|V'|}{V}.
\end{equation}
Without the Gauss-Bonnet term, if $\mathcal{SC}$II is satisfied,
then it is impossible to satisfy $\mathcal{SC}$I for single field slow-roll inflation with $\Delta N\sim 60$.
With the help the Gauss-Bonnet term,  it is very easy to satisfy both $\mathcal{SC}$I and $\mathcal{SC}$II conditions,
as long as $1-\lambda$ small enough, for example $1-\lambda<\Delta N^{-1}c^{-1}$.

In summary, with the help of the Gauss-Bonnet term, as long as the order one parameter $\lambda$ is close to 1,
the two swampland criteria $\mathcal{SC}$I and $\mathcal{SC}$II can be easily satisfied,
and the tensor to scalar ratio $r$ is also consistent with the observations \cite{Akrami:2018odb}.
From Eq. \eqref{ns1}, we see that the parameter $\lambda$ have no effect on the scalar spectral tilt $n_s$, so the constraint on $n_s$ can also be satisfied.

In the next section, we will use two inflationary models, the power-law potentials and the E-model, as examples to support the above discussion.

\section{The models}
\label{sec2}
In the following we consider two inflation models, the chaotic inflation model and the E-model.
We show that with the help of the Gauss-Bonnet term, the two swampland criteria \eqref{swampland1} and \eqref{swampland2} are satisfied for both models.
Additionally, the models also satisfy the observational constraints \cite{Akrami:2018odb},
\begin{equation}\label{obsers}
  n_s=0.9649\pm0.0042,\quad r_{0.002}<0.064.
\end{equation}

\subsection{The power-law potential}

\begin{figure}[htbp]
  \centering
  \includegraphics[width=0.45\textwidth]{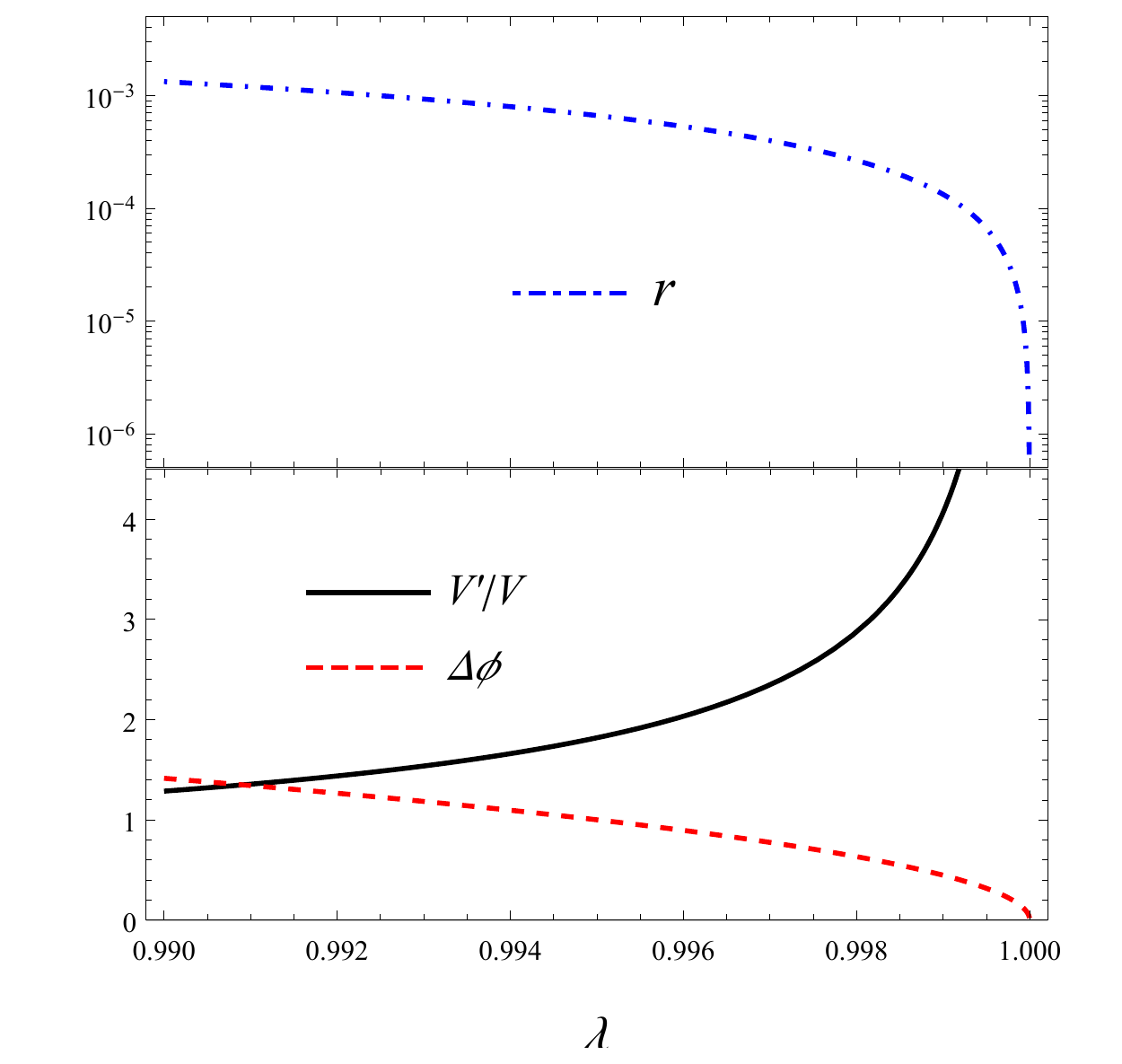}
  \caption{The dependence on $\lambda$ for the power-law potential with $p=2$.
  The upper panel shows the tensor to scalar $r$. The lower panel shows the gradient of the potential $V'/V$ and the field excursion $\Delta \phi$.}
  \label{pchao}
\end{figure}

For the chaotic inflation with the power-law potential \cite{Linde:1983gd}
\begin{equation}\label{chap}
  V=V_0\phi^p,
\end{equation}
the excursion of the inflaton  is
\begin{equation}\label{chaphin}
  \Delta \phi=\sqrt{2(1-\lambda)p}\left(\sqrt{N+\tilde{n}}-\sqrt{\tilde{n}}\right),
\end{equation}
where $N$ is the remaining number of  $e$-folds before the end of inflation, and
\begin{equation}\label{chani}
\tilde{n}=
\left\{
  \begin{array}{cc}
   p/4, & 0<p<2, \\
 (p-1)/2,& p\geq2. \\
  \end{array}
\right.
\end{equation}
The scalar spectral tilt $n_s$ and the tensor to scalar ratio $r$ are
\begin{gather}
\label{chaons}
  n_s-1=-\frac{p+2}{2(N+\tilde{n})},\\
\label{chaor}
  r=\frac{4(1-\lambda)p}{N+\tilde{n}}.
\end{gather}
From Eq. \eqref{chaons}, we see that $n_s$ is independent on  $\lambda$.
If we choose $p=2$ and $N=60$, the scalar spectral tilt is $n_s=0.9669$, which is consistent with the observations \eqref{obsers}.
By varying the value of $\lambda$, the values of the gradient of the potential $V'/V$,
inflaton excursion $ \Delta \phi$ and tensor to scalar ratio $r$ are shown in Fig. \ref{pchao}.
As the value of $1-\lambda$ becomes smaller and smaller, $r$ and $\Delta \phi$ will become smaller and smaller too,
but $V'/V$ will become larger and larger. As long as $1-\lambda$ is small enough,
the two swampland criteria \eqref{swampland1} and \eqref{swampland2} as well as the observational constraints \eqref{obsers} are satisfied.
For example, if we chose $1-\lambda=5\times 10^{-5}$, we have
\begin{gather}
\label{chaons1}
n_s=0.9669,\quad r=6.6\times 10^{-6},\\
\frac{V'}{V}=18.2,\quad \Delta \phi=0.1.
\end{gather}
The predictions \eqref{chaons1} are consistent with the observations \eqref{obsers}, and both the field excursion $\Delta \phi$ and
the gradient of the potential $V'/V$ satisfy the two swampland criteria.

\subsection{The E-model}
\begin{figure}[htbp]
  \centering
  \includegraphics[width=0.45\textwidth]{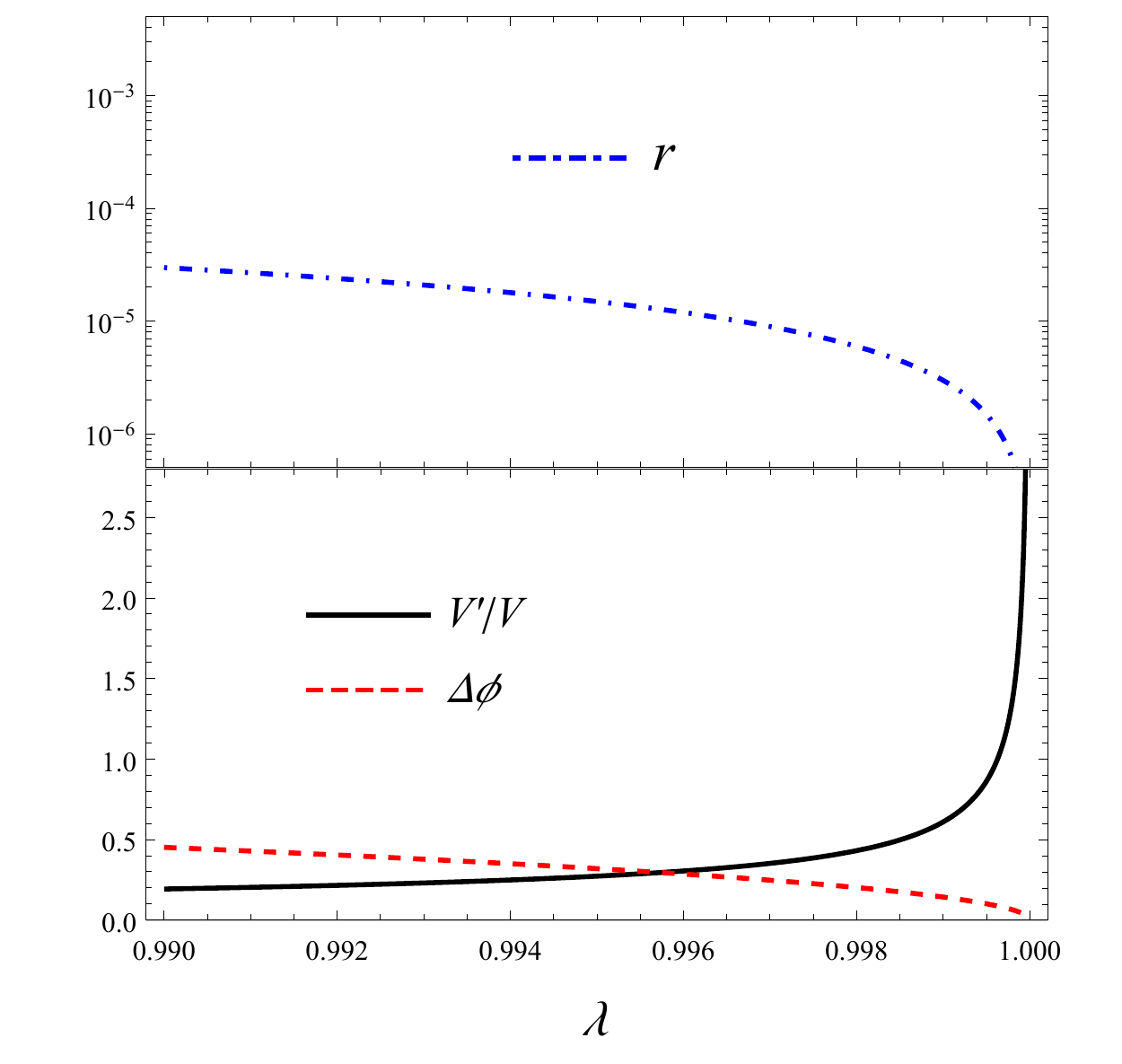}
  \caption{The dependence on $\lambda$ for the the E-model with $\alpha=1-\lambda$.
  The upper panel shows the tensor to scalar $r$. The lower panel shows the gradient of the potential $V'/V$ and the field excursion $\Delta \phi$.}
  \label{pemodel}
\end{figure}

For the E-model \cite{Kallosh:2013maa,Carrasco:2015rva}
\begin{equation}\label{epotential}
  V=V_0\left[1-\exp\left(-\sqrt{\frac{2}{3\alpha}}\phi\right)\right]^{2n},
\end{equation}
the excursion of the inflaton for $n=1$ is
\begin{equation}\label{ephi}
\Delta \phi=-\frac{1}{\sqrt{6\alpha}}\left[3\alpha(\tilde{N}+X-1)+4(1-\lambda)N\right],
\end{equation}
where
\begin{equation}\label{en}
\tilde{N}=1+W_{-1}\left[-X\exp\left(-X-\frac{4(1-\lambda) N}{3\alpha}\right)\right],
\end{equation}
and $X=2\sqrt{1-\lambda}/(\sqrt{3\alpha})+1$, the  function $W_{-1}$ is the lower branch of the Lambert W function.
The scalar spectral tilt $n_s$ and the tensor to scalar ratio $r$ are \cite{Yi:2016jqr}
\begin{gather}
\label{ens}
  n_s=1+\frac{8(1-\lambda)}{3\alpha\tilde{N}}-\frac{16(1-\lambda)}{3\alpha\tilde{N}^2}, \\
  r=\frac{64(1-\lambda)^2}{3\alpha\tilde{N}^2}.
\end{gather}
If we chose $\alpha=1-\lambda$ and $N=60$, we get $n_s=0.9678$ which is consistent with the observation.
Varying $\lambda$, the values of $V'/V$, $ \Delta \phi$ and $r$ are shown in Fig. \ref{pemodel}.
Similar to the chaotic inflation, as the $1-\lambda$ becomes smaller and smaller, $r$ and $\Delta \phi$ become smaller and smaller,
and $V'/V$ become larger and larger. As long as $1-\lambda$ is small enough,
the two swampland criteria \eqref{swampland1} and \eqref{swampland2} as well as the observational constraints \eqref{obsers} are satisfied.
If we chose $1-\lambda=10^{-4}$, we get
\begin{gather}
\label{ens1}
n_s=0.9678,\quad r=3.0\times 10^{-7},\\
\frac{V'}{V}=1.9,\quad \Delta \phi=0.045.
\end{gather}
The predictions \eqref{ens1} are consistent with the observations \eqref{obsers}, and both the field excursion $\Delta \phi$ and
the gradient of the potential $V'/V$ satisfy the two swampland criteria.

\section{conclusion}
\label{sec3}

The two swampland criteria pose threat on the single filed slow-roll inflation.
With the help of the Gauss-Bonnet coupling, the relationship between the $r$ and $V'/V$ is described by Eq. \eqref{r2},
i.e., $r$ is reduced by a factor of $(1-\lambda)^2$ compared with the result in standard single field slow-roll inflation.
Because of the reduction in $r$, the first swampland criterion is easily satisfied by requiring $r$ to be small. On the other hand,
it is easy to satisfy the second swampland criterion by requiring $1-\lambda$ to be small and keeping $r$ to be small.

For the chaotic inflation with $p=2$, if we take $1-\lambda=5\times 10^{-5}$, we get
$n_s=0.9669$, $r=6.6\times 10^{-6}$, $V'/V=18.2$ and $\Delta \phi=0.1$.
Therefore, for the chaotic inflation with $p=2$ and $\lambda>0.99995$,
the models satisfy not only the observational constraints, but also the two swampland criteria.
 For the E-model, If we chose $1-\lambda=10^{-4}$, we get
$n_s=0.9678$, $r=3.0\times 10^{-7}$, $V'/V=1.9$ and $\Delta \phi=0.045$.
The model satisfies not only the observational constraints, but also the two swampland criteria.
In conclusion, the Gauss-Bonnet inflation with the condition \eqref{xiphi} satisfies not only the observational constraints, but also the swampland criteria.
Therefore, the swampland criteria may imply the existence of Gauss-Bonnet coupling.

\begin{acknowledgments}
This work was supported in part by the National Natural Science
Foundation of China under Grant Nos. 11875136 and 11475065 and the Major Program of the National Natural Science Foundation of China under Grant No. 11690021.
\end{acknowledgments}

%

\end{document}